\documentclass[a4paper,11pt]{article}
\pdfoutput=1 
\usepackage{jinstpub} 
\usepackage[section]{placeins}
\RequirePackage{graphicx}
\RequirePackage{flushend}
\usepackage[style=numeric,sorting=none]{biblatex}
\usepackage{lineno}
\usepackage{ulem}
\usepackage{amsmath}

\title{Low Energy Electronic Recoils and Single Electron Detection with a Liquid Xenon Proportional Scintillation Counter}

\author[1]{Jianyang~Qi,%
\note{Email: jiq019@ucsd.edu}}
\author{Noah Hood,}
\author[2]{Abigail Kopec,%
\note{Email: akopec@physics.ucsd.edu}}
\author{Yue Ma, Haiwen Xu, Min Zhong,}
\author[3]{Kaixuan~Ni%
\note{Email: nikx@physics.ucsd.edu}}

\affiliation{Department of Physics, University of California San Diego, La Jolla, CA, 92093, USA}

\abstract{Liquid xenon (LXe) is a well-studied detector medium to search for rare events in dark matter and neutrino physics. Two-phase xenon time projection chambers (TPCs) can detect electronic and nuclear recoils with energy down to kilo-electron volts (keV). In this paper, we characterize the response of a single-phase liquid xenon proportional scintillation counter \mbox{(LXePSC)}, which produces electroluminescence directly in the liquid, to detect electronic recoils at low energies. Our design uses a thin (10--25~$\mu$m diameter), central anode wire in a cylindrical LXe target where ionization electrons, created from radiation particles, drift radially towards the anode, and electroluminescence is produced. Both the primary scintillation (S1) and electroluminescence (S2) are detected by photomultiplier tubes (PMTs) surrounding the LXe target. Up to 17~photons are produced per electron, obtained with a 10~$\mu$m diameter anode wire, allowing for the highly efficient detection of electronic recoils from beta decays of a tritium source down to $\sim$1~keV. Single electrons, from photoemission of the cathode wires, are observed at a gain of 1.8~photoelectrons (PE) per electron. The delayed signals following the S2 signals are dominated by single-photon-like hits, without evidence for electron signals observed in the two-phase xenon TPCs. We discuss the potential application of such a LXePSC for reactor neutrino detection via Coherent Elastic Neutrino Nucleus Scattering (CE$\nu$NS). }


\begin{document}
\maketitle
\flushbottom

\section{Introduction}
\label{sec:intro}

Dual-phase Liquid Xenon Time Projection Chambers (LXeTPCs) have traditionally been used in large scale rare event searches, and operate by detecting the prompt scintillation light (S1) and the proportional electroluminescence of ionization electrons (S2) from an energy deposition. However, these detectors have never achieved perfect charge collection efficiency in practice~\cite{XENON:2022ltv,LZ:2022ufs}. Additionally, they display a background comprised of delayed single electrons that can last $\mathcal{O}(1)$~s after a large S2 signal~\cite{XENONCollaborationSS:2021sgk,LUX:2020vbj,Kopec:2021ccm}. This background can impact low energy event searches which are only capable of producing S2s, such as those from Coherent Elastic Neutrino Nucleus Scattering (CE$\nu$NS)~\cite{Ni:2021mwa}. Two main hypotheses for this background are that there are electrons trapped on impurities and then are released, or trapped at the liquid gas interface and are extracted later than most of the S2 electrons. This begs the question of whether or not it is possible to make a detector with a sensitivity to single electrons, which does not have a liquid-gas interface. Proportional scintillation in liquid xenon was first demonstrated in 1979 by Masuda et al~\cite{Masuda:1978tjp} and was proposed for dark matter searches in the 1990s~\cite{Park:1994hd, Wang:1998gq}. More recent work includes a working detector constructed at Columbia in 2014~\cite{Aprile:2014ila}, and a simulation study for the prospects of few-electron signal detection~\cite{Kuger:2021sxn}. In addition to investigating whether or not the origin of the delayed electron background is due to a liquid-gas interface, a single-phase design has the potential to improve the LXeTPCs by eliminating the liquid-gas interface that is currently preventing the major LXe-based dark matter detectors to achieve perfect electron extraction efficiency. The aforementioned papers use a design similar to the dual-phase TPC where the electric field is oriented along $\hat{z}$ in a cylindrical volume. In this paper, we use a cylindrical detector design with a radial field, proposed in~\cite{Lin:2021izy}, which we name the Liquid Xenon Proportional Scintillation Counter (LXePSC). With this technology, we were able to estimate the average light collection efficiency and ionization gain (photoelectrons detected per electron), also known as $g_1$ and $g_2$, respectively, for a variety of anode voltages. In addition, we were able to observe low-energy electronic recoils from tritium beta decays. Single electrons, emitted via the photoelectric effect from the cathode wires, are observed based on their distinct timing and spectrum. However, with low amplification, an unambiguous single electron waveform is not yet identifiable due to their similar pulse shape to the spurious light emission.

Previously, we have tested such a detector using a 25~$\mu$m diameter anode wire to produce electroluminescence~\cite{Wei:2021nuk}. In this paper, we improve and further characterize the performance using a 10~$\mu$m diameter anode wire. The updates to our previous LXePSC design (formerly called the radial TPC) is described in section~\ref{sec:tpc}. The detector performance is described in section~\ref{sec:performance}. In section~\ref{sec:cs137}, we show the light detection efficiency ($g_1$) and ionization gain ($g_2$) for different anode-to-cathode voltages as obtained from the $^{137}$Cs calibration. The tritium calibration including the low energy electronic recoil band is described in~\ref{sec:tritium}. We also discuss our investigation into the signals after a large S2 in section~\ref{sec:single_electrons} to show evidence of single electron signals. Lastly, we discuss the major issue of light emission at high fields in section~\ref{sec:lightemission}, and the future work needed to improve the detector design.

\section{Liquid Xenon Proportional Scintillation Counter (LXePSC)}
\label{sec:tpc}

\subsection{LXePSC with different anode wire diameters}
\label{subsec:design}
\begin{figure}[htbp]
    \centering 
    \includegraphics[height = 0.28\textheight]{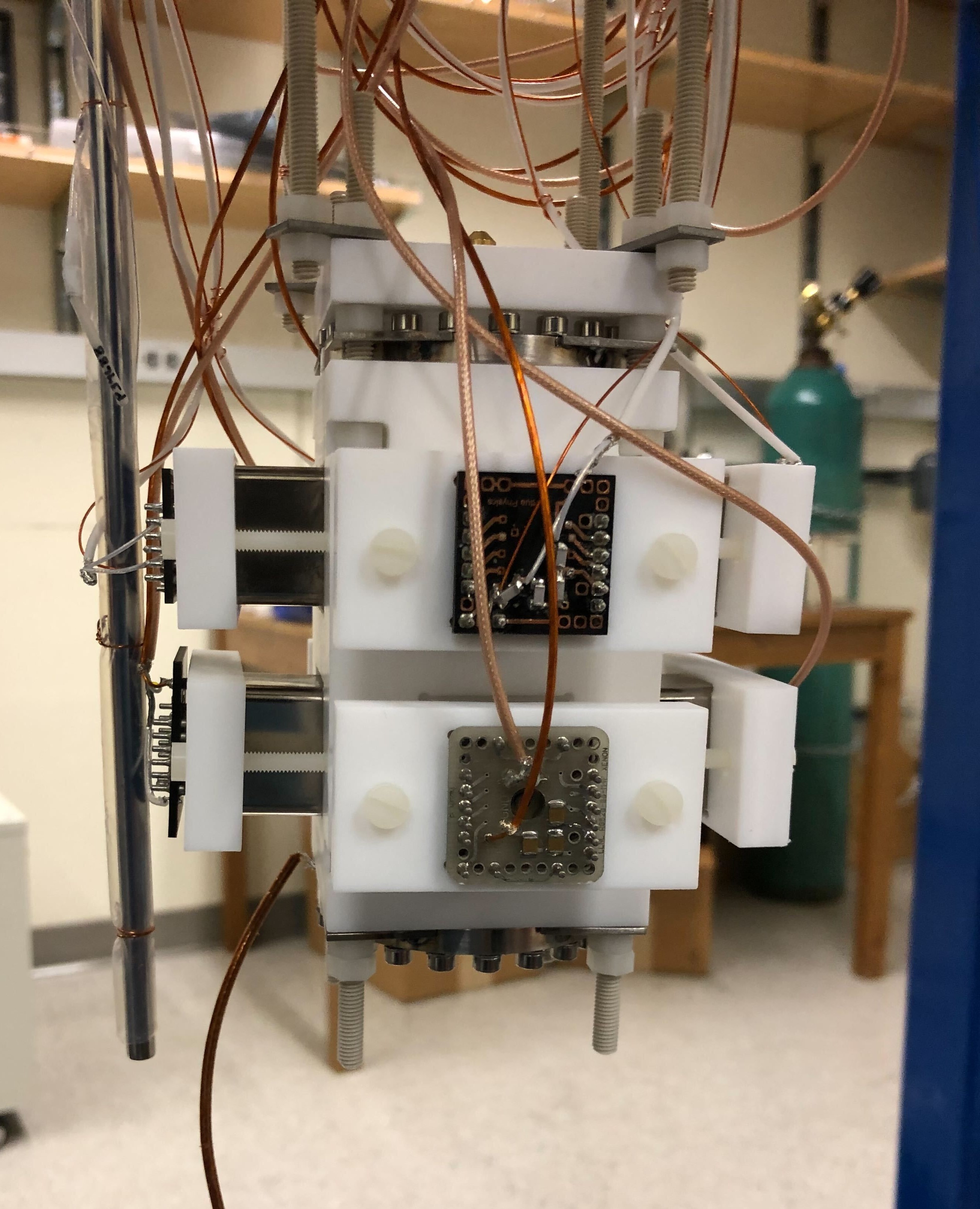}
    \includegraphics[height = 0.3\textheight]{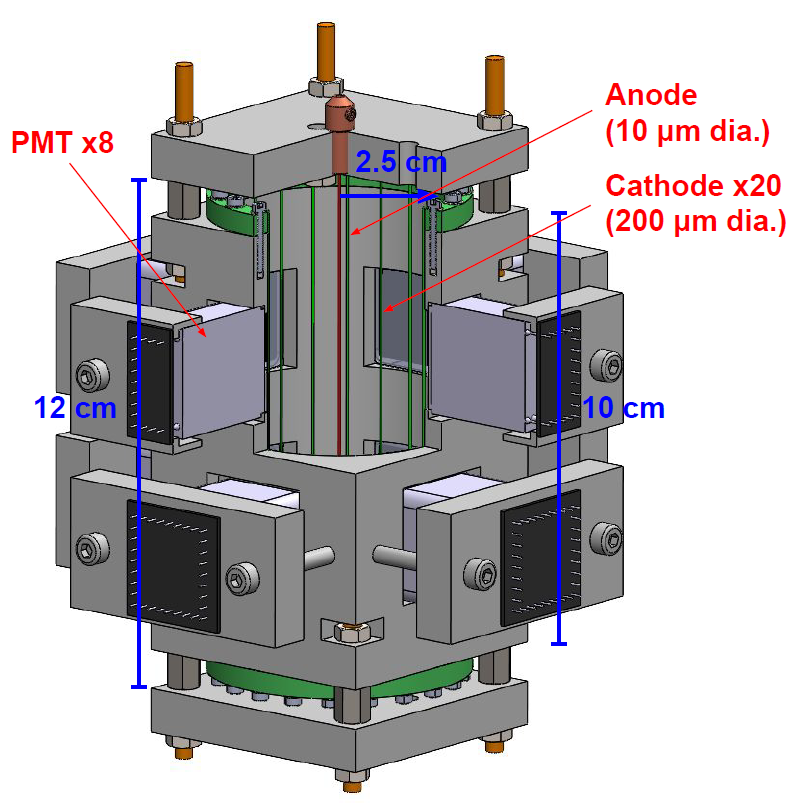}
    \caption{\textbf{Left}: A picture of the LXePSC during the installation of the 10~$\mu$m diameter anode. \textbf{Right:} The design drawing of the detector. The wire diameters are scaled up so that they are visible.}
    \label{fig:detector_drawings}
\end{figure}

The design of the Liquid Xenon Proportional Scintillation Counter (LXePSC) is a cylindrical LXe target where the S2 is produced near an anode wire in the center of the detector. Twenty cathode wires are at the edge of the sensitive LXe target and simultaneously act as a shield for the Photomultiplier Tubes (PMTs). This design is similar to that discussed in a previous conference proceeding with a 25~$\mu$m diameter anode wire~\cite{Wei:2021nuk}, except for two adjustments. The first is that we replaced the L-shaped anode holder at the top and bottom of the detector (see Fig. 1 of~\cite{Wei:2021nuk}) with Accu-Glass Push-On 0.040 connectors. The second is that we replaced the 25~$\mu$m diameter anode wire with a 10~$\mu$m diameter  wire (both are gold-plated tungsten wires from California Fine Wire Co.). This is to investigate whether we can get a higher gain of photons per electron, as we can achieve a larger maximum electric field with a thinner wire given a fixed anode-to-cathode voltage. These modifications can be seen in Fig. \ref{fig:detector_drawings}.

\subsection{Operation and data taking}
\label{subsec:operation}

\begin{figure}
    \centering
    \includegraphics[width = 0.49\textwidth]{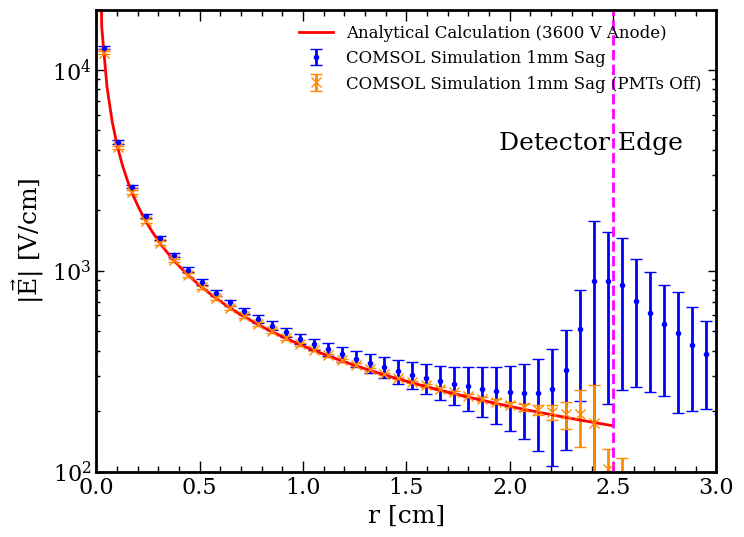}
    \includegraphics[width = 0.48\textwidth]{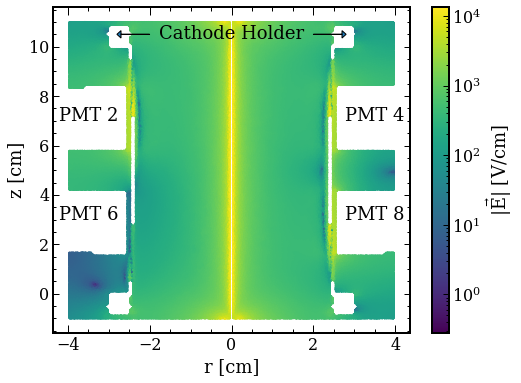}
    \caption{\textbf{Left:} The electric field as a function of $r$ for the central 2.47~cm (in $z$) of the detector. The error bars refer to the standard deviation of the electric field in $z$ and $\theta$ at a particular $r$. \textbf{Right:} A map of the magnitude of the electric field for a slice along the $r-z$ plane. The electric field was simulated using COMSOL\textsuperscript{\texttrademark} multiphysics for both figures.}
    \label{fig:field_sims}
\end{figure}

During operation, we noticed that one of the PMTs (PMT 6) was shorted to the cathode. This was later discovered to be caused by a stray piece of 10~$\mu$m wire, which connected the bottom cathode ring to the PMT. As a result, we needed to turn off this PMT and ground the cathode. This led to some field non-uniformity around PMT 6 as the other PMTs were set to voltages around -700~V to -600~V. The PMT voltages were set such that each PMT had a gain of $10^6$~e$^-$/photoelectron (PE). This was calibrated by using a pulse-generator to drive a green LED and trigger our digitizer externally. The resultant area spectrum was then fit with four gaussians to account for the noise, single PE, double PE, and triple PE spectra, and the gain was taken to be the mean of the single PE distribution. This procedure was repeated for different voltages, and a power-law fit to the gain as a function of PMT voltages gave us our nominal PMT voltages for a gain of $10^6$~e$^-$/PE. Furthermore, the negative voltage of the PMTs and the grounded cathode led to some leakage of the PMTs' potential into the drift field region. At the end of the run, we opened the detector and noticed that there was some inward bending of the cathode wires. This is modeled in the simulation by assuming that the cathode wires are parabolas that bend inward towards the anode. The "sag" of a cathode wire is referred to as the maximum displacement of this parabola from the edge of the inner cylinder of the detector. To characterize the field, we simulated the case where the sagging is 1~mm, 2~mm, and 3~mm. Our field simulations are summarized in Fig.~\ref{fig:field_sims}. This simulation assumes a 3600~V anode, grounded cathode, and PMTs at their operating voltages. The simulated electric field from COMSOL is larger than the analytic calculation, due to the effect of the negative high voltage PMTs being near the grounded cathode. When the PMTs are grounded, this effect goes away and the field is more consistent with the analytical calculation. Despite these difficulties, we were still able to see a clear $^{137}$Cs photopeak, as well as a low energy electronic recoil (ER) band from tritium decays.

\begin{figure}[!htbp]
    \centering
    \includegraphics[width = 0.9\textwidth]{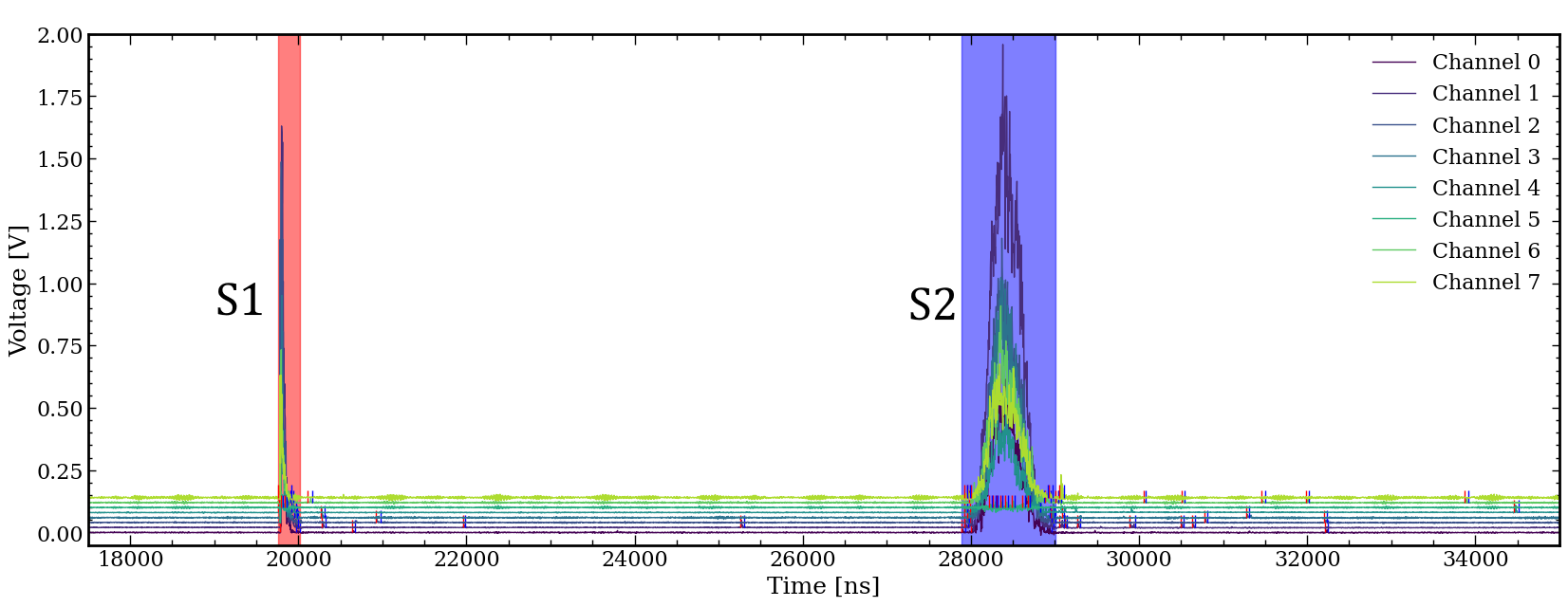}
    \caption{Example waveform for a $^{137}$Cs event for 3~kV anode and grounded cathode. S1 highlighted in red, S2 highlighted in blue.}
    \label{fig:cs137_waveform}
\end{figure}

We used a CAEN V1720 digitizer which took full waveforms with a trigger, and an event window between 50~$\mu$s and 1~ms depending on which data type is being taken. One of the primary issues with our previous work~\cite{Wei:2021nuk} was peak-finding with high levels of light emission. This peak-finding used to be done on the waveform summed across all channels, which is much noisier than the per-channel waveforms. To fix this, we selected only the \textit{per-channel} pulse hits that were at least 2 standard deviations from the baseline noise. The pulse hits which overlap across channels are then summed into \textit{peaks}. In this way, the noise from channels which do not see a pulse hit is not added. Similar to the two-phase xenon TPCs,  the S1 and S2 signal have different waveform shapes in the LXePSC. The S1 shape looks like the sum of two decaying exponentials due to de-excitation, and the S2 shape looks more-or-less gaussian due to the diffusion of electrons. As such, the rise-time is defined as the time it takes for the peak to traverse from 10\% of its maximum height to 90\% of its maximum height. S1s are classified as having a rise-time of less than 40~ns, and S2s have a rise-time greater than 40~ns. The S1s and S2s are then grouped into \textit{events} (see Fig. \ref{fig:cs137_waveform} for an example). This is done by looking at each S1 in a trigger window, starting from the largest S1 by area, and finding the biggest S2 within 20~$\mu$s after the S1. The second largest S1 or S2 found in this window (if any exist) are considered the alternate S1 or S2.

\section{Performance of the LXePSC}
\label{sec:performance}
\subsection{Detector calibration with $^{137}$Cs}
\label{sec:cs137}
To calibrate the response of the detector, we need to see the S1 and S2 response to a monoenergetic source. We used a $^{137}$Cs 661.7~keV gamma source placed in a cup on the outer vessel of the detector, and vertically aligned near the middle of the detector in $z$. Previously, we achieved $g_1 = 0.13$~PE/$\gamma$ and $g_2 = 0.7$ PE/e$^-$ with all eight PMTs turned on, and with a 4~kV anode and -750~V cathode; this corresponds to an anode-surface field of 495~kV/cm~\cite{Wei:2021nuk}. However, the $^{137}$Cs photopeak of our previous run was significantly smeared in (S1, S2), especially at anode voltages greater than 4-kV, which prevented us from measuring $g_1$ and $g_2$ for a variety of anode voltages. In this run, we achieved a higher $g_2$ of $1.6\pm 0.2$~PE/e$^-$ at an anode voltage of 3.6~kV despite one of the PMTs being off. Considering $g_1$, this value of 1.6~PE/e$^-$ corresponds to $\sim 17$ photons produced per electron by electroluminescence in the liquid via the analysis in section \ref{subsec:g1g2_av}. It is important to note that, in this context, $g_2$ is referred to as the \textit{ionization gain} rather than the single electron gain (SE gain). Although there is no incomplete extraction, we expect a small effect from the electron-lifetime, so the size of the S2 from the $^{137}$Cs peak is only comprised of the ionization electrons which did not attach to impurities. As such, $g_2$ is an underestimation of the SE gain. Furthermore, we were able to measure the $g_1$ and $g_2$ values across multiple anode voltages. At above 3.6~kV on the anode, we start to observe spurious light emission as discussed in section~\ref{sec:lightemission}.

\subsubsection{Selection of the $^{137}$Cs photo-absorption events}
\label{sec:cs137_cuts}
To select for the full deposited 661.7~keV, we need to first cut out multiple scatters. This is done using two cuts. The first cut compares the main S1 or S2 to the alternate S1 or S2. If the alternate S1 (S2) area is comparable in size to the main S1 (S2) area, then it is likely due to a multiple scatter event. The cut threshold is set such that the alternate S2 (S1) area of a single scatter is less than 2\% (10\%) of the main S2 (S1) area. This cut keeps 49\% of the original data. The second cut deals with the case that the multiple scatters happen close to each other, in which case the diffusion of the electrons from each scatter will merge the two S2s into one peak. These are cut using the Jenks natural breaks algorithm from the \textit{strax} data processor used in XENONnT~\cite{strax}, which gives a "goodness-of-split" score (GOS) ranging from 0 to 1. Multiple scatters of this type have a GOS>0.75, and this cut keeps 87\% of the S2 peaks.



\begin{figure}[!htbp]
    \centering
    \includegraphics[height = 0.29\textheight]{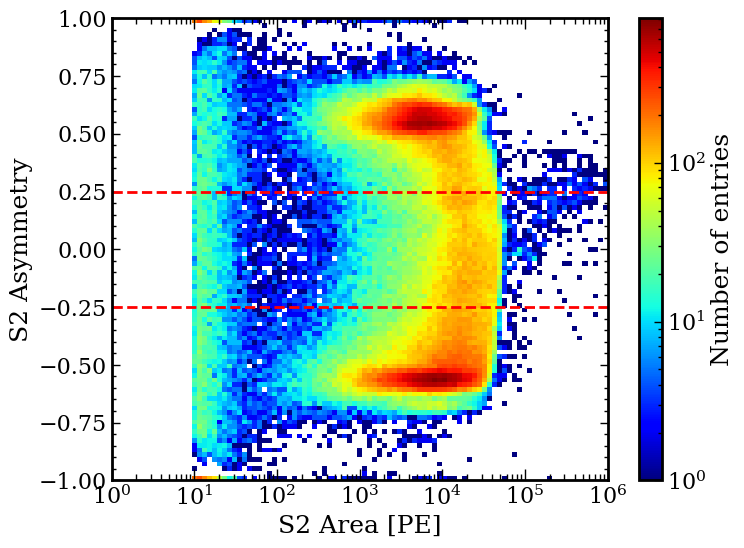}
    \includegraphics[height = 0.29\textheight]{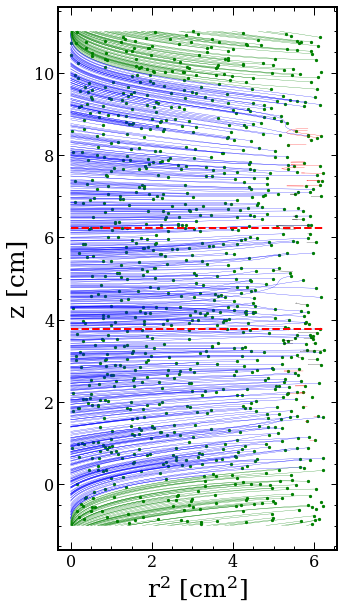}
    \caption{\textbf{Left:} Asymmetry vs S2 area distribution for the $^{137}$Cs dataset. \textbf{Right:} Electron cloud tracks through $r$ and $z$. Blue paths successfully reach the anode, green paths stop at the wall, and red paths reach a cathode wire. The electrons clearly drift from their initial positions (denoted by green dots) towards the top or bottom plate. The top and bottom red dashed lines are the estimated $z$ positions which correspond to an S2 asymmetry of -0.25 and 0.25 respectively. The electron paths are generated using the \textit{cylinterp} code \cite{cylinterp}.}
    \label{fig:z_cut}
\end{figure}

Events near the top and bottom PTFE plates are also cut, as this region has a difficult-to-model electric field that affects the charge yield and detection efficiency, and significant radioactive backgrounds from materials. Furthermore, events near the top and bottom of the detector tend to have S2 electrons which follow field lines with considerable drift in $\hat{z}$, meaning that the initial $z$ position of the event is not the same $z$ position at which the S2 light is produced (Fig. \ref{fig:z_cut} right). Therefore, we select for events near the center of the detector in $z$. To do this, we use the S2 asymmetry, defined as \begin{equation}
    S2_{asym} = \frac{S2_{\text{top 4 PMTs}}-S2_{\text{bottom 4 PMTs}}}{S2_{\text{top 4 PMTs}}+S2_{\text{bottom 4 PMTs}}}
 \end{equation} as a proxy for $z$, and select for events between $S2_{asym}\in [-0.25, 0.25]$. Here, $S2_{\text{bottom/top 4 PMTs}}$ is the \textit{integrated} area across the bottom (top) four PMTs. There are two ways to estimate the $z$-range that corresponds to this asymmetry cut ($z_{selection}$). The first is to look at the tritium events (Sec.~\ref{sec:tritium}), which should be uniformly distributed in $z$. The ratio of tritium events with $S2_{asym}\in [-0.25, 0.25]$ to the total number of tritium events is equal to $z_{selection}/12$~cm. Here 12~cm is the height of the active volume. From this, we can estimate that $z_{selection}$ corresponds to the central 2.47~cm of the detector. However, this estimation has the underlying assumption that all events are in a region where the ionization electrons can drift to the anode and produce an S2 (a charge sensitive volume), which may not be true. To check this issue, we also ran an optical simulation in GEANT4 \cite{GEANT4:2002zbu}, where there is no such charge insensitive volume that decreases the total number of measured events. We assumed a PTFE reflectivity of 99\%, and generated $10^5$ photons along $10^4$ randomly sampled positions near the anode wire, and computed the asymmetry for each simulated position, which gave us a $z_{selection} = 2$~cm. However, the difference between $z_{selection} = 2$~cm or $z_{selection} = 2.47$~cm only yields a 1\% difference in the average electric field, with the latter yielding a higher field variability. As such, we use $z_{selection} = 2.47$~cm to be conservative in the estimation of the systematic uncertainty in $g_1$ and $g_2$.

It should be noted that since PMT 6 was off during this run, the calculation of the S2 asymmetry also did not include the area seen by PMT 2 (the PMT right above PMT 6). The asymmetry parameter loses $z$ resolution near the top and bottom of the detector, and starts to saturate, as seen in Fig. \ref{fig:z_cut} left.

\begin{figure}
    \centering
    \includegraphics[width = 1\textwidth]{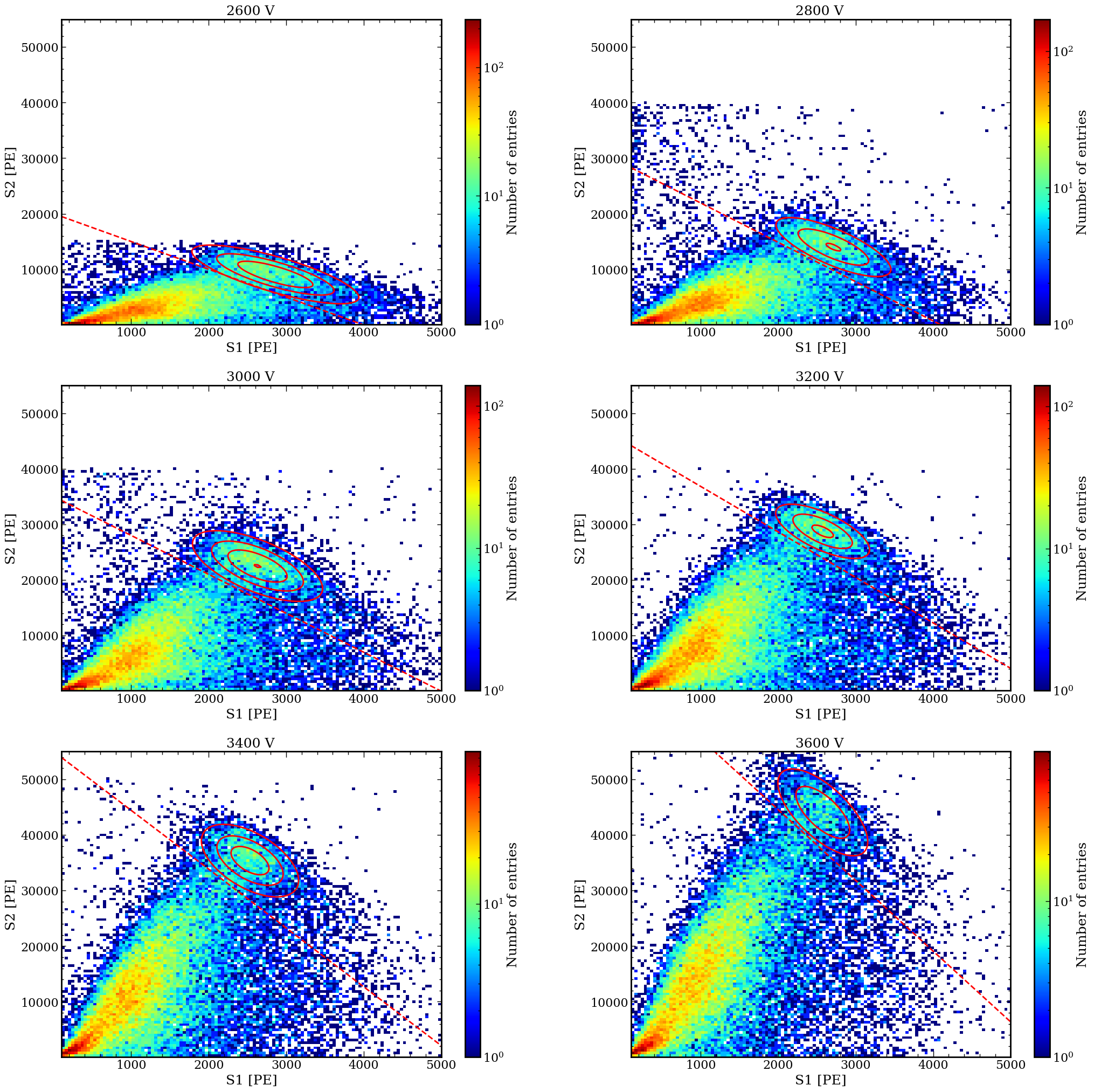}
    \caption{$^{137}$Cs S2 vs S1 distribution for different anode voltages from 2600 to 3600 V. The S1 is suppressed while the S2 is enhanced as the anode voltage increases. The red contours are for the 2D Gaussian fit corresponding to the 661.7-keV photo-absorption peak, with the dashed, red cut-line.}
    \label{fig:Cs137_voltage_sweep}
\end{figure}

After the $z$ and multiple scatter cuts are applied, we can see the S1 and S2 distribution shows a population of high S1 and high S2 which is anti-correlated, this is the 661.7~keV $^{137}$Cs photopeak. As we increase the anode voltage, we see that the S2 signal is enhanced while the S1 is suppressed, as shown in Fig.~\ref{fig:Cs137_voltage_sweep}. We select for the photopeak by visually inspecting the end of the Compton shelf in $(S1, S2)$ space and drawing a line through it, then selecting for the events in $(S1, S2)$ that lie above that line. Afterwards, we fit the population using a 2-D gaussian to find the center value for S1 and S2, referred to as $S1_c$ and $S2_c$. There is an ambiguity in choosing the cut line, so we use two possible selections for the photopeak by using two different cut lines. The first cut line is determined by eye to estimate the beginning of the photopeak events and the end of the Compton shelf. The second cut line is determined by varying the first line's S2-intercept by five percent. Together, these two measurements for $S1_c$ and $S2_c$ are then used in determining the systematic uncertainty of $g_1$ and $g_2$.

\subsubsection{Extracting $g_1$ and $g_2$ from $^{137}$Cs data}
\label{sec:cs137_sims}

\begin{figure}[!htbp]
    \centering
    \includegraphics[width = 0.9\textwidth]{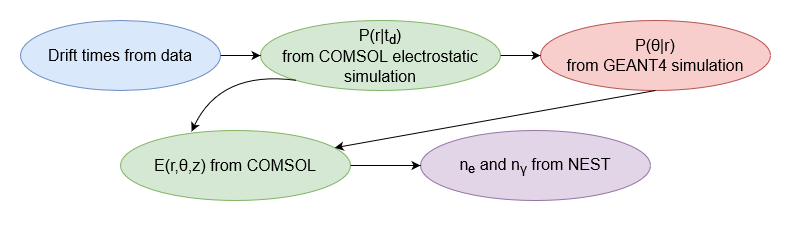}
    \caption{The simulation chain used to get the absolute number of photons and electrons from the $^{137}$Cs peak.}
    \label{fig:mc_chain}
\end{figure}

In dual-phase LXeTPCs, $g_1$ and $g_2$ are estimated via calibrations using multiple monoenergetic sources. However, we only had the $^{137}$Cs source available to us at this time, so we estimate our $g_1$ and $g_2$ by finding $g_1 = S1_c / \langle n_\gamma \rangle$ and $g_2 = S2_c / \langle n_e \rangle$. Here, $n_\gamma$ and $n_e$ are the number of S1 (S2) photons (electrons) given by the Noble Element Simulation Technique (NEST) \cite{szydagis_m_2018_1314669}, which takes the electric field and deposited energy as input. The deposited energy is simply 661.7~keV, however, to find the electric field, we must first get the position of the events, followed by a multi-step simulation chain as summarized in Fig.~\ref{fig:mc_chain}.

\begin{figure}[!htbp]
    \centering
    \includegraphics[width = 0.53\textwidth]{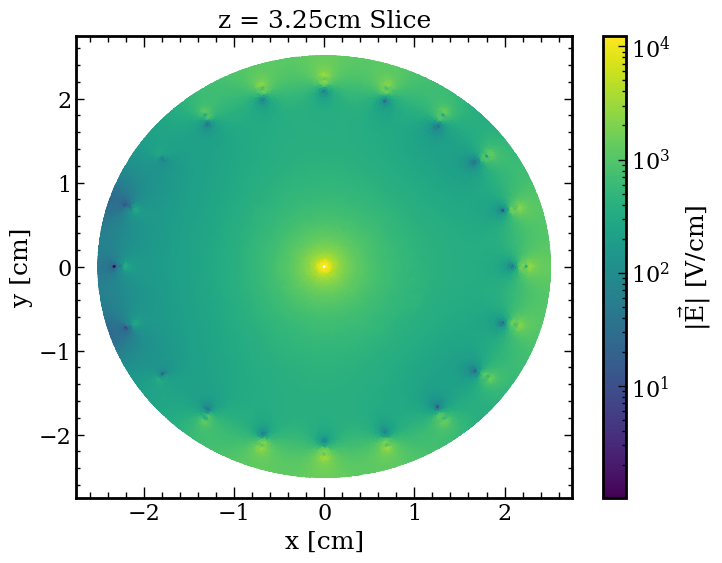}
    \includegraphics[width = 0.4\textwidth]{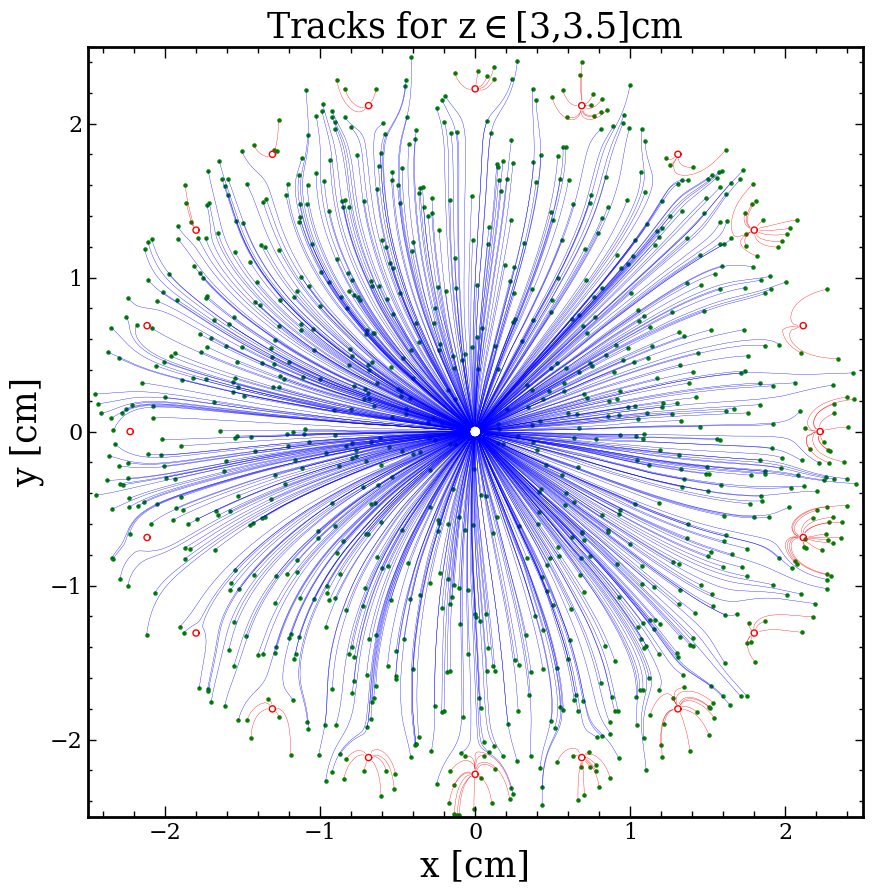}
    \caption{\textbf{Left:} Field map at z = 3.25~cm. The effect of PMT 6 being off is clear, as the electric field in the left side of the detector for a $z$ slice going through this PMT is much lower. This is due to the aforementioned issues discussed in Sec.~\ref{subsec:operation}. In this simulation, we also assumed that the cathodes sagged by 3mm at the middle of the detector \textbf{Right:} Tracks of electrons. Slices in drift time, $t_d$, correspond to different $r$ distributions, $P(r|t_d)$. For smaller drift times, there is a near one-to-one correspondence between $r$ and $t_d$, however, the bending of the field lines for events near the edge of the detector smear this correspondence for large $t_d$. The red paths correspond to electrons which hit the cathode due to the electric field between the negative PMTs and the grounded cathode. Since PMT 6 is grounded, we do not see any red paths near this PMT. This also leads to a charge-insensitive volume.}
    \label{fig:field_dt}
\end{figure}

\begin{figure}[!htbp]
    \centering
    \includegraphics[width = 0.49\textwidth]{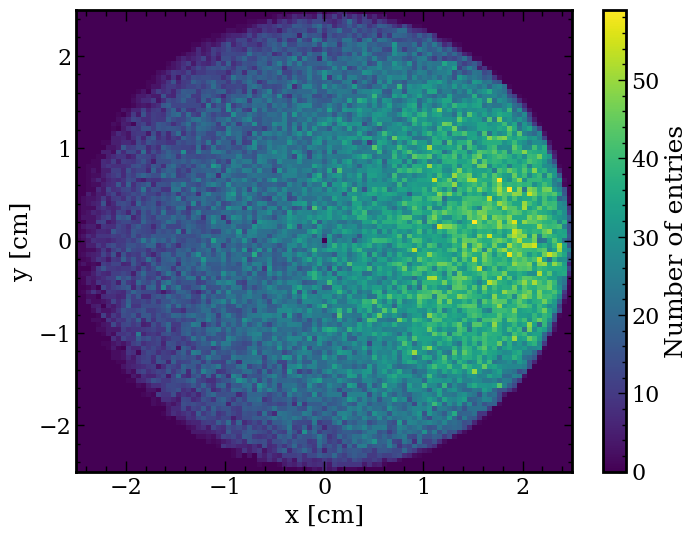}
    \includegraphics[width = 0.49\textwidth]{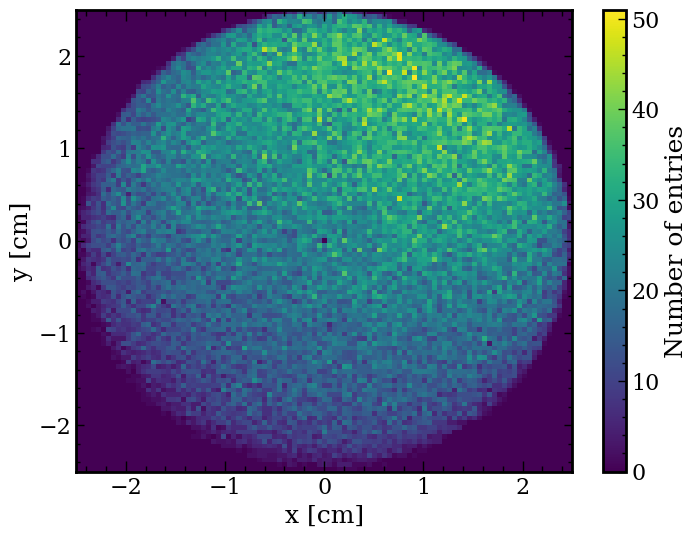}
    \caption{GEANT4 $^{137}$Cs photopeak distributions with a source placed at different locations.}
    \label{fig:g4_cs137}
\end{figure}

While our radial ($r$) position can be inferred from the drift time ($t_d$) via an electric field simulation using COMSOL (see Fig.~\ref{fig:field_dt}), our $z$ and $\theta$ positions are not properly reconstructed. For $z$, the asymmetry cut lets us sample from the central 2.47~cm of the detector. This, coupled with the fact that our source is placed in the middle of the detector, lets us uniformly sample $z$ within this range. For the $\theta$ distribution, we first use GEANT4 \cite{GEANT4:2002zbu} to simulate the position distribution of $^{137}$Cs events within the detector. This simulation gives us $P(\theta | r)$ (see Fig. \ref{fig:g4_cs137}), where $r$ is reconstructed from the drift time. However, there is an ambiguity of where the source is placed with respect to the detector in $\theta$. This comes from the fact that we did not precisely measure the orientation of the source position with respect to the orientation of the detector. As such, we did three GEANT4 simulations based on three different possible positions that the $^{137}$Cs source could be. Lastly, we simulated the electric field assuming either no cathode wire sag, or a cathode wire sag of 3~mm (see Sec.~\ref{subsec:operation}). For each simulation, we obtained $n_\gamma$ and $n_e$. The two possible choices of the data photopeak selection, the three possible choices of the $^{137}$Cs source position, and the two possible values of cathode wire sag gives twelve possible values of $g_1$ and $g_2$, for which a systematic uncertainty can be calculated. All of these effects contribute a 2.4\% uncertainty in $g_1$ and 2.7\% for $g_2$ at a 3.6~kV anode. However, the largest systematic uncertainty comes from the NEST light and charge yields. According to Fig.~2 of \cite{Szydagis:2022ikv}, this corresponds to approximately 6~photons/keV for the number of photons and 6~electrons/keV for the number of electrons. The NEST yield uncertainties contribute a 19\% and 14\% uncertainty to $g_1$ and $g_2$ at 3.6~kV, respectively. The statistical uncertainty is negligible.

\subsubsection{$g_1$ and $g_2$ as a function of anode voltage}
\label{subsec:g1g2_av}

\begin{figure}[!htbp]
    \centering
    \includegraphics[width = 1\textwidth]{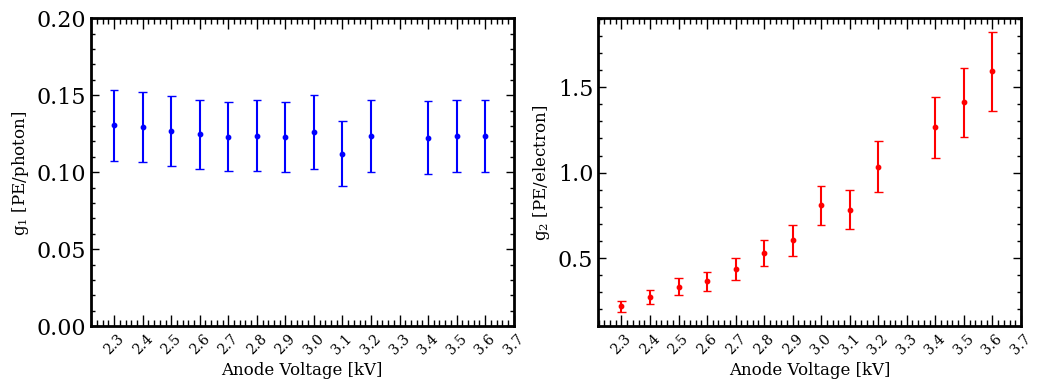}
    \caption{$g_1$ and $g_2$ for the values across different anode voltages. The systematic uncertainty comes from the possible positions of the $^{137}$Cs source, the photopeak selection in S2 vs S1 space, and the light and charge yield uncertainties in NEST \cite{Szydagis:2022ikv}.}
    \label{fig:g1g2}
\end{figure}

The $g_1$ and $g_2$ values obtained from the above data selection and simulation are shown in Fig.~\ref{fig:g1g2}. $g_1$ is approximately constant while $g_2$ increases with the anode voltage, as expected. Here, since there is no gas gap and all electrons not lost to electronegative impurities are collected, $g_2$ serves as a slight underestimate to the effective SE gain, accounting for the purity. The "effective" label is due to the fact that the electric field around the anode is greater than the charge-multiplication threshold of 725~kV/cm, according to the model proposed by Aprile et al.~\cite{Aprile:2014ila} (henceforth referred to as the Columbia model). As such, the total light seen will be due to \textit{both} the primary electron as well as secondary electrons produced via charge-multiplication. However, this effect is small, we estimate that about 4\% of the light produced is due to the secondary electrons, according to the Columbia model.

\begin{figure}[!htbp]
    \centering
    \includegraphics[width = 0.6\textwidth]{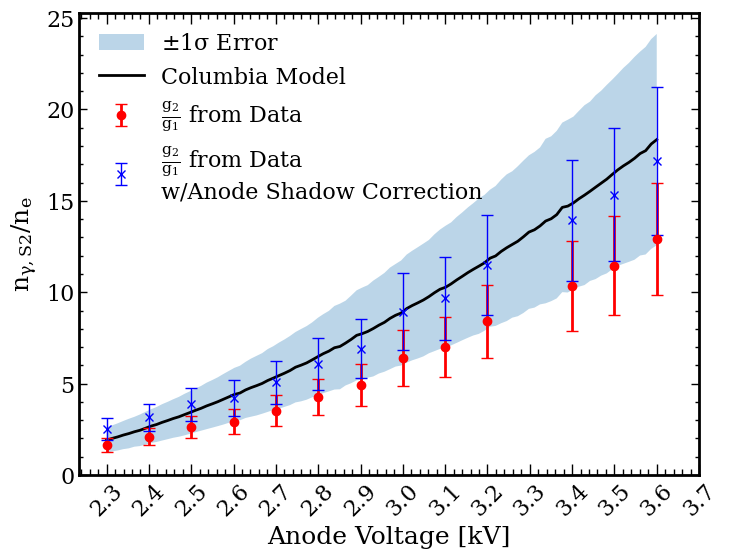}
    \caption{A comparison of electroluminescence yield $n_\gamma/n_e$ to the Columbia model~\cite{Aprile:2014ila}. We see that up to 17$\pm$4~photons are produced per electron at an anode voltage of 3.6~kV. The upper limit of $n_\gamma/n_e$ is given according to Eq.~\ref{eq:anode_shadow}. The 1$\sigma$ shaded region is computed by sampling the parameters given in the Columbia model with their associated uncertainties.}
    \label{fig:anode_shadow_ngamne}
\end{figure}

It is also possible, in principle, to estimate the absolute (pre-efficiency) number of photons produced per electron (electroluminescence yield) by looking at the ratio $g_2/g_1$. Ideally, \begin{equation}
    g_2 = (n_{\gamma,S2}/n_e)\langle LCE(S2) \rangle P_{det} \hspace{1cm} g_1 = \langle LCE(S1) \rangle P_{det}.
\end{equation} Here, $n_{\gamma,S2}$ is the number of photons produced by the S2 electrons near the anode, not the number of photons from S1. $P_{det}$ is the detection probability and is a property of the PMTs which will cancel out in $g_2/g_1$. $\langle LCE(S1) \rangle$ refers to the light collection efficiency (LCE) averaged over the event positions, and $\langle LCE(S2) \rangle$ refers to the light collection efficiency averaged over the region around the anode which can produce electroluminescence. If we use the Columbia model, then the field threshold for electroluminescence is 412~kV/cm, meaning that the electroluminescence threshold radius, $r_T$, for our setup with a 3.6~kV anode is around 10~$\mu$m. This means that the anode may block a considerable amount of light from an S2. To estimate this effect, consider a source of light is at a distance $r$ from the anode with radius $r_a$. The probability that a photon will hit the anode is $P_{hit} = \arcsin(r_a/r)/\pi$. As such, the probability that a photon will either avoid the anode or reflect off of it is \begin{equation}
    P_{escape}(r) = (1-P_{hit}) + P_{hit}\rho_{gold}
\end{equation} where $\rho_{gold}$ is the reflectivity of gold. We can average this over $r\in[r_a, r_T]$ to get $P_{escape} = \langle P_{escape}(r) \rangle$. This can give us an upper limit on $n_\gamma/n_e$ by naively assuming that $\langle LCE(S2) \rangle \approx \langle LCE(S1) \rangle P_{escape}$, and $\rho_{gold} = 0$. Which gives \begin{equation}
    \frac{n_{\gamma, S2}}{n_e}\bigg\vert_{\text{upper limit}} \approx \frac{g_2}{g_1}\frac{1}{\langle 1 - \arcsin(r_a/r)/\pi \rangle}.
    \label{eq:anode_shadow}
\end{equation} We call this effect the "anode shadowing correction", and our $n_\gamma/n_e$ is consistent with the Columbia model as shown in Fig.~\ref{fig:anode_shadow_ngamne}. At the maximum anode voltage of 3.6~kV, the $g_2$ of 1.6~PE/e$^-$ corresponds to 17$\pm$4~photons produced by an electron.

\subsection{Low energy electronic recoils from tritium beta decays}
\label{sec:tritium}

\begin{figure}[!htbp]
    \centering
    \includegraphics[width = 0.9\textwidth]{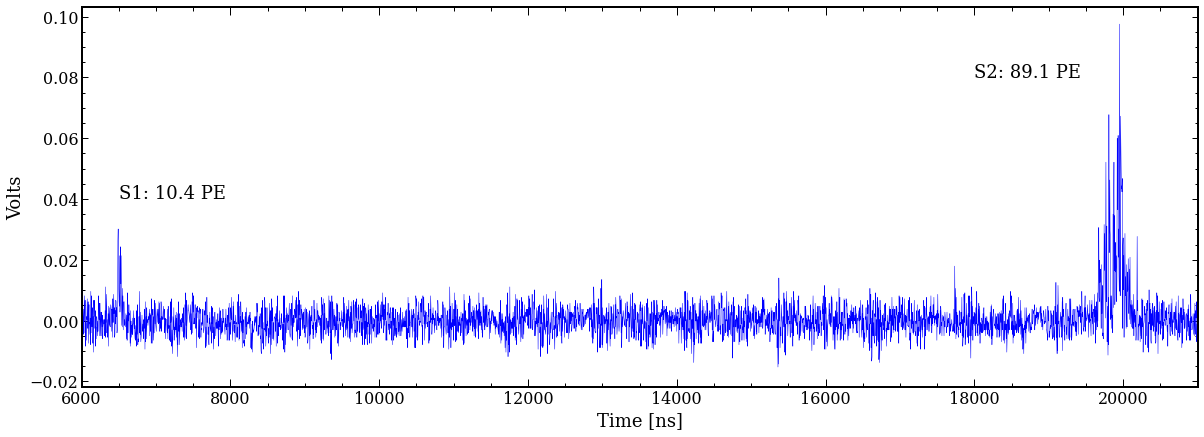}
    \caption{Waveform from a tritium event.}
    \label{fig:trit_wf}
\end{figure}

We injected a tritiated methane source into our detector in order to probe the response to low energy electron recoils (ER) of a few keV. These events populate a band in $(S1, \log(S2/S1))$ space (Fig.~\ref{fig:tritium_energy_band}), and we can count the rate of these events in the band as a function of time. During this calibration, the anode was at a voltage of 3600~V. The tritiated methane was removed from the detector by the SAES getter in the gas circulation loop over the course of three days.

\begin{figure}[!htbp]
    \centering
    \includegraphics[width = 0.49\textwidth]{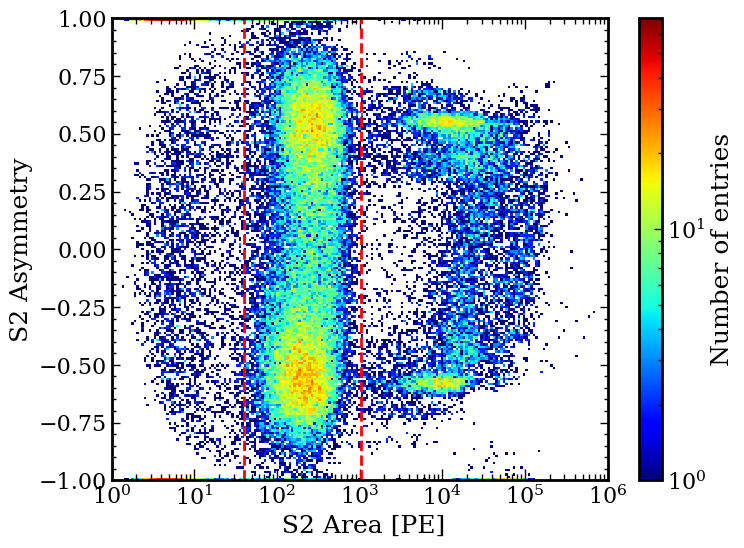}
    \includegraphics[width = 0.49\textwidth]{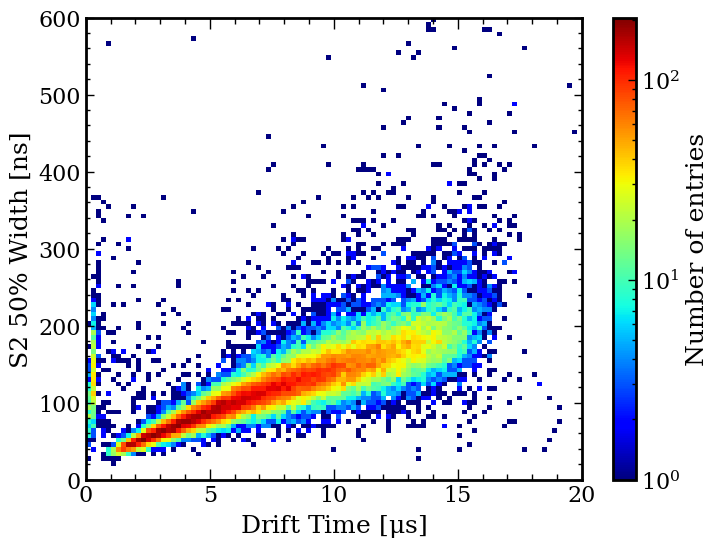}
    \caption{\textbf{Left:} The asymmetry vs S2 area distribution after applying the multiple scatter cuts. The tritium events are selected between the red lines. \textbf{Right:} The tritium events' S2 50\% width vs drift time.}
    \label{fig:tritium_physicality}
\end{figure}

\begin{figure}
    \centering
    \includegraphics[width = 0.9\textwidth]{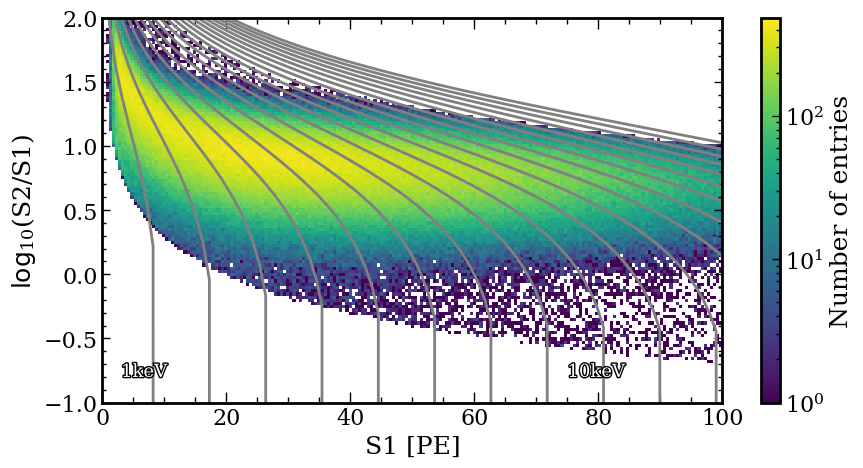}
    \caption{The low energy ER band from the tritium events after applying the aforementioned cuts. The gray lines are energy contours in 1~keV, with energy reconstruction discussed in the text.}
    \label{fig:tritium_energy_band}
\end{figure}

To select for the tritium events, we use the same multiple scatter and $z$ cuts as explained in section~\ref{sec:cs137_cuts}. An example of a tritium event waveform is shown in Fig.~\ref{fig:trit_wf}. We can easily see the population of tritium events by looking at the S2 areas as shown in Fig.~\ref{fig:tritium_physicality}. Tritium events are easily identifiable (Fig.~\ref{fig:tritium_physicality}) as their S2 size is significantly higher than any accidental-coincidences, yet significantly lower than higher-energy detector backgrounds. The charge yield for tritium events is less affected by electric field variations near the wall, since they are lower energy compared to $^{137}$Cs events, according to NEST~\cite{szydagis_m_2018_1314669}. Therefore, the tritium S2 areas do not show significant variation with extremes in asymmetry and indicate uniform light collection efficiency all along the anode. As explored later, this indicates that the average number of measured electroluminescence photons are a constant in $z$ for all electrons that reach the anode. The greater spread in absolute asymmetry values at lower energies corresponding to the walls, is due to lower total photon counts distributed on the top and bottom PMTs. Furthermore, we can confirm the physicality of these events by observing that their S2 50\% width increases with the drift time. Here, the 50\% width is defined as the time between the 25\% and 75\% area percentiles of the S2 waveform, and diffusion causes the S2 waveforms corresponding to larger drift times to be wider. For these events, we can use the $g_1$ and $g_2$ values from the $^{137}$Cs calibration to draw energy contours in $(S1, \log(S2/S1))$ space by computing $E = W(S1/g_1 + S2/g_2)$. Here, $W = 13.5$~eV from the NEST fit of $W$ as a function of density, here we used a density of 2.89~g/cm$^3$~\cite{Szydagis:2022ikv}. This low energy electronic recoil band with energy contours is shown in Fig.~\ref{fig:tritium_energy_band}.

The explicitly non-uniform, approximately $1/r$, electric field causes the number of electrons (photons) for a given energy deposition to decrease (increase) as the $r$ location of the event increases due to a greater recombination of electrons in weaker electric fields~\cite{Doke:2002oab}. Thus, the S2 area should decrease with increasing $r$. However, the inverse relationship between S2 area and $r$ can also be caused by the attachment of electrons to electronegative impurities~\cite{XENON:2019ykp}. In dual-phase TPCs, this is corrected by multiplying the S2 by the factor $e^{t_d/\tau_e}$, where $\tau_e$ is known as the electron lifetime, and can be calibrated by measuring the S2 response using a mono-energetic source~\cite{XENON:2019izt}. In LXePSCs, we would need an external purity monitor to give us this information, as we cannot disentangle the effects of attachment to impurities from the decrease in charge yield due to the suppressed electric field. It is worth mentioning that this effect may also be present in the dual-phase TPC, as XENON1T showed that there is a discrepancy in the electron lifetime when using different calibration sources~\cite{XENON:2019ykp}. However, the discrepancy is around 10\%, while the LXePSC has much larger field variations.

To correct S1 and S2 properly, we would need to have $g_1$ and $g_2$ as functions of position. In principle, this could be done by calibrating $g_1$ and $g_2$ for each selection of $r$ (i.e. drift time). However, this would require more statistics than we currently have.

\subsection{Delayed signals and photoemission electrons}
\label{sec:single_electrons}

The lowest achievable energy threshold is set by the smallest detectable S2, which is the electroluminescence of a single electron. However, previous two-phase liquid xenon dark matter experiments have observed higher rates of single and few electron signals following a bright interaction in the detector~\cite{XENON100:2013wdu,LUX:2020vbj,Kopec:2021ccm,XENONCollaborationSS:2021sgk,Bodnia:2021flk}. Within one maximum drift time duration, electrons released by a bright S2's photons drift to the anode and are measured. Considering material quantum efficiencies and geometric effects, photon-induced single electron rates are estimated to be on the order of $10^{-4}$~e$^-$/PE in the LXePSC, based on observations by XENON100~\cite{XENON100:2013wdu}. In two-phase detectors, rates of single electron signals remain elevated after a maximum drift time, decreasing according to a power law for up to a second. This implies that these electrons cannot be from prompt photoemission~\cite{LUX:2020vbj,XENONCollaborationSS:2021sgk}. The two major hypotheses for the high single and few electron signal rates after a maximum drift time are that they were once ionization electrons from the interaction that were either captured on and later released from impurities in the xenon bulk, or stuck at and later released from the liquid xenon surface~\cite{Kopec:2021ccm}. The LXePSC would be able to rule out the liquid surface hypothesis if there are delayed electrons.

However, in addition to delayed electrons, XENON1T and LUX observed elevated rates of single-photon-like signals (lone hits) following an interaction event that decay with a similar power law~\cite{LUX:2020vbj,XENONCollaborationSS:2021sgk}. The lone hit rates were roughly an order of magnitude higher than the single electron signal rates in XENON1T, but the opposite is indicated in LUX. Without a liquid surface, the LXePSC expects to measure all electrons released in the detector, unless they are captured on electronegative impurities or drift to the wall. The maximum drift time of the detector is relatively short compared to the electron lifetime governing the capture of electrons on electronegative impurities, so most electrons in an interaction in the charge-sensitive region reach the anode. Therefore, the $g_2$ in the center of the detector is a good indicator (although slight underestimate) of the SE gain, or the number of photons measured per electron at the anode. At this time, a single electron signal would predominantly appear as one or two photons, and any observed power law of rates requires a signal spectrum analysis to determine whether it would be more characteristic of single electrons or photon lone hits. 

To explore delayed signals, background data without a calibration source was taken with 1~ms event windows and the anode at 3.6~kV. Despite a higher particle interaction rate in the detector, the digitizer was only able to handle a 10~Hz trigger rate to save 1ms event windows, thus saving only 10~ms of detector livetime per second. Electronic recoil events were selected based on S1 area, S2 area, and S2 width considering diffusion and drift time between the S1 and S2. Only a lower bound was set for the S2 area of $10^3$~PE, in order to encompass all bright electronic recoils from cosmogenic particles and intrinsic detector radioactivity. The largest S2 signals exceeded $10^5$~PE. Background data was necessary to capture the longest possible windows between interactions. The asymmetry cut was not applied because it would have cut 91\% of valid electronic recoil interactions in the charge-sensitive region, which are concentrated near the walls. Although variations in the electric field near the walls affect the charge yield, particularly for higher energy events, the calibration with low-energy tritium beta decay events does not show a significant variation in measured S2 area across the detector and validates the assumption that the SE gain is constant for all electrons that reach the anode. Cuts against multiple scatters, as discussed previously, were applied. These selected events were bright enough to explore photoemission, and are good candidates to explore delayed signals. After each event, all signals (including lone hits and peaks of two or more hits) were analyzed. Because these were single-scatter events, the signals were smaller than 10~PE. The $g_2$ measured in Sec.~\ref{subsec:g1g2_av} for these detector conditions was 1.6~PE/e$^-$, so lone hit signals confined to one PMT would be typical of both signals of delayed electrons reaching the anode and photons. However, the signal rates of peaks with two contributing PMTs would have a similar magnitude to the lone hits if delayed electrons dominated, or the rates would be consistent with lone hit pile-up if photons dominated. An analysis of the signals' spectra is therefore required. 

\begin{figure}[!htbp]
    \centering
    \includegraphics[height = 0.3\textheight]{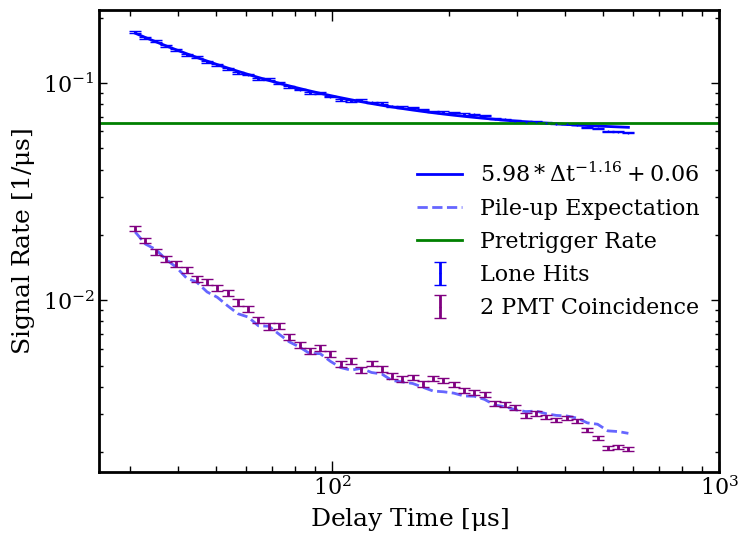}
    \caption{Rates of photon-like signals in single PMTs (lone hits) and rates of peak signals with coincident photon-like hits in two PMTs. The two-PMT peaks are consistent with pile-up of lone hits. The pretrigger lone hit rate (expected to be dominated by intrinsic PMT dark counts) is also indicated.}
    \label{fig:photon_power}
\end{figure}

The delayed signals were available for all events in the 1~ms windows from two maximum drift times (30~$\mu$s) up to 600~$\mu$s. The two maximum drift time requirement before analyzing delayed signals was chosen to avoid contamination from photoemission electron signals. The rates of signals that are lone hits in single PMTs and are peaks comprised of two coincident hits in two PMTs are shown in Fig.~\ref{fig:photon_power}. The lone hit rate before the S1 is also included in the figure as the horizontal green line, which is expected to be dominated by intrinsic PMT dark counts. The configuration of the digitizer created significant dead time, so information of particle interactions just before the captured event window is unknown. Since the pretrigger window is 200~$\mu$s, it is likely that delayed signals from previous events during dead time contaminate this pretrigger window. Therefore, the pretrigger lone hit rate is higher than the asymptotic rate of the delayed lone hit rates at the end of the window above about 400~$\mu$s. The lone hit rate decay is consistent with a power law plus a constant, as observed in LUX and XENON1T~\cite{LUX:2020vbj,XENONCollaborationSS:2021sgk}. The power-law power is steeper than observed in XENON1T, with a power of -1.16 compared to -0.7. We do not address the absolute rate amplitudes, which would need to be adjusted by detector electron survival efficiencies and normalized by the previous S2 areas. The expected pile-up of lone hits is consistent with the rates of signals with two coincident PMTs, indicating no significant contribution from single electrons.

\begin{figure}[!htbp]
    \centering
    \includegraphics[height = 0.23\textheight]{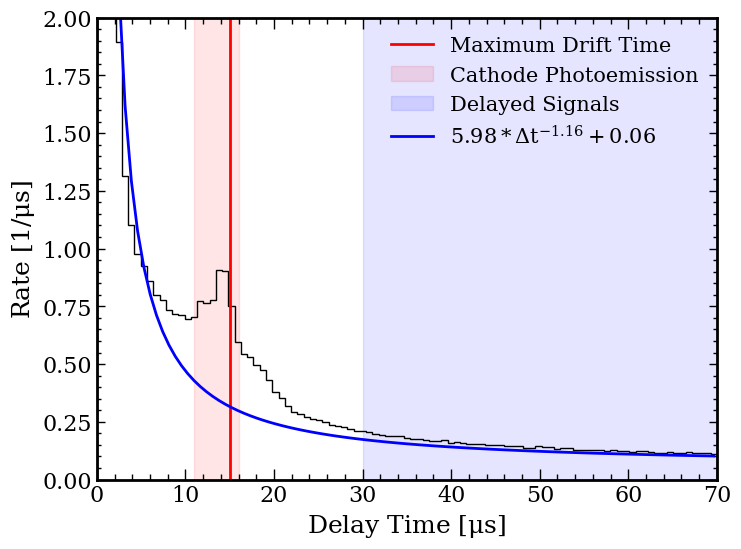}
    \includegraphics[height = 0.23\textheight]{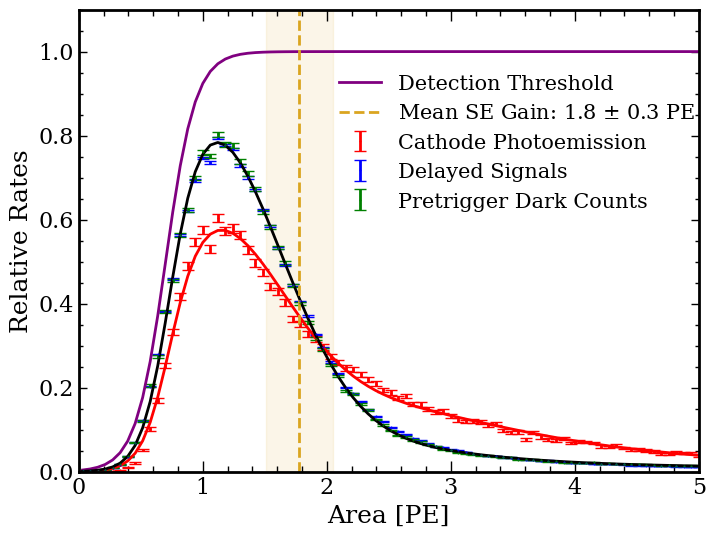}
    \caption{\textbf{Left:} Signal rates with time after s2s. A peak in signals that is attributed to electrons emitted from the cathode is shaded in red, with a red vertical line at the maximum drift time. Delayed signals are shaded blue. The power law from Fig.~\ref{fig:photon_power} is shown as a blue line and estimates a 44.2\% contamination to the pure photoemission electron signals in the shaded red region. \textbf{Right:} The data points outlining the area spectra of these corresponding red and blue signal populations are shown, along with the detection threshold (purple line), and mean area of the cathode photoemission electron signals (vertical gold dashed line with shaded uncertainty). Lone hits that occur before the S1 in the event are added as green data points, and their spectrum is fit with a black line. The photoemission spectrum fit is marked with a red line.}
    \label{fig:delayed_signals}
\end{figure}

To better differentiate if the delayed signals are more electron-like or photon-like, we analyzed the spectra. The pretrigger signals are dominated by dark count lone hits and are single-photon-like. S2s are bright enough that we expect some prompt photoemission electrons. The highest probability for photoemission is from the metal cathode wires, so higher rates of single electrons are measured exactly one maximum drift time after a bright S2. The left panel of Fig.~\ref{fig:delayed_signals} shows the rates of all signals after the good events, with the maximum drift time of 15~$\mu$s denoted with a vertical red line. The power law from Fig.~\ref{fig:photon_power} was extrapolated (blue curve in the left panel of Fig.~\ref{fig:delayed_signals}) to determine the contamination from delayed signals.  The photoemission electron population is taken from the red shaded region, and the delayed signal contamination in that region under the blue curve is estimated to be 44.2\%. 

The right plot in Fig~\ref{fig:delayed_signals} shows the signal area spectra for the cathode photoemission signals, and delayed signals. The spectrum of signals from before the S1 in the event (expected to be dominated by PMT dark counts) are also included for reference. The spectra were fit with a series of gaussians with the $n$th peak having a mean of $n$ and standard deviation $\sqrt{n}$ larger than the 1~PE peak. The delayed signals are consistent with dark counts, whereas the photoemission is significantly different. Subtracting the 44.2\% contamination of the delayed signal spectrum from the photoemission spectrum, the expectation value for pure photoemission single electrons (SE gain) is 1.8$\pm$0.3~PE. This is consistent with the $g_2$ at this anode voltage of 1.6~PE/e$^-$ and indicates that these signals are characteristic of single electrons. Without needing to account for extraction efficiency and with a sufficient purity for a small detector, the $g_2$ is expected to be a slight underestimation of the true SE gain accounting for a small fraction of electrons lost on electronegative impurities. After iterative fitting, the photoemission spectrum constrains both the delayed signals and pretrigger signals to contain less than 5\% single electron signals. 

Based on the spectrum, we can also conclude that the delayed signals and pretrigger dark counts are not dominated by xenon scintillation photons. Xenon photons on our PMT photocathodes have sufficient energy to cause two photoelectrons about 20\% of the time~\cite{Faham:2015kqa}. Both LUX and XENON1T also observed that delayed photons were inconsistent with xenon scintillation~\cite{LUX:2020vbj,XENONCollaborationSS:2021sgk}. The delayed photons are hypothesized to be lower energy photons in the visible range, potentially due to Teflon fluorescence. However, they have not been ruled out as an intrinsic PMT dark count background: they may be elevated thermal photocathode emission following exposure to more than $\mathcal{O}(1000)$~photons. The pretrigger and delayed signal spectra observed in our data are consistent with the previous experiments. Less than 10\% of signals are larger than 1~PE. The pretrigger photon lone hits are explored in more detail in the following sections.

Our observation of signals after an event led to a confirmed observation of single electrons with a gain of 1.8~PE/e$^-$. This is the first confirmed observation of single electron electroluminescence in liquid xenon, although a distinct single electron signal waveform is still elusive. Our observation takes advantage of photoemission by the cathode wires to select the most pure sample of single electron signals. While we do observe elevated signal rates that decay with a power law for long times after an interaction, these signals are more consistent with photon backgrounds and cannot be dominated by single electrons.

\section{Light Emission and Discussion}
\label{sec:lightemission}

\begin{figure}[!htbp]
    \centering
    \includegraphics[width = 0.9\textwidth]{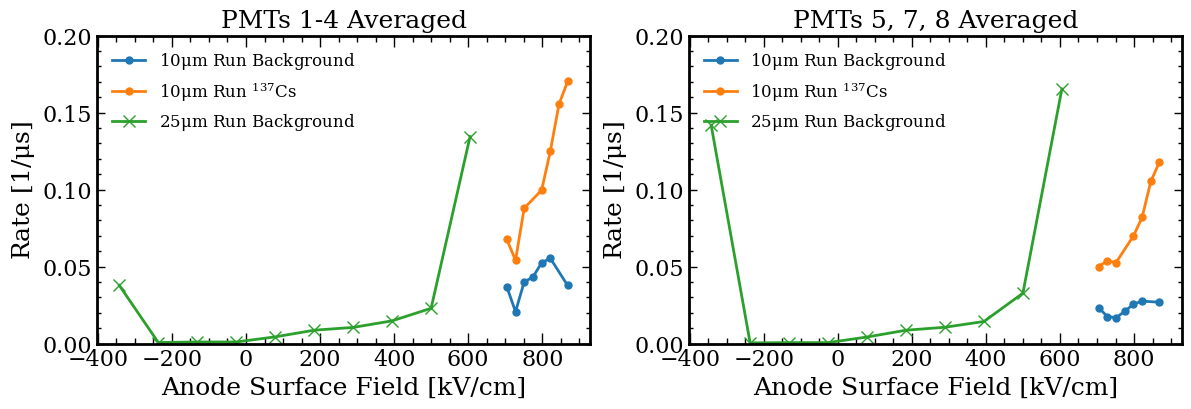}
    \caption{Rates of pulse hits occurring before the main S1 in the event window averaged over each row of PMTs. The "anode surface field" is given by $\Delta V/\ln(r_c/r_a)r_a$, where $r_c$ ($r_a$) is the cathode (anode) radius, and $\Delta V$ is the anode to cathode voltage.}
    \label{fig:lightemission}
\end{figure}

The biggest limitation to the single-phase design is the low $g_2$ value. While one may attempt to ramp-up the anode voltage indefinitely, this also increases the rate of spurious light emission, which we saw in our previous work~\cite{Wei:2021nuk}. This light emission, dominated with lone hit signals, acts as a constant noise to our baseline, for which a pile-up of single photons could obfuscate a single electron signal. In this run, we were able to obtain a higher electric field around the anode before observing an unmanageable level of light emission (Fig.~\ref{fig:lightemission}). The light emission rate is calculated by counting the number of pulse hits within the first 10~$\mu$s of a 100~$\mu$s event window. Pulse hits found within this region occur before the triggering S1. We also saw that the light emission rate is higher for data taken with a $^{137}$Cs source near the detector, compared with background data. This suggests that the light emission rate is related to the event rate in the detector.

The optimum $g_2$ value will require a detailed balance between the anode diameter and electric field. Assuming there exists some sort of minimum electric field, $E_{min}$, for which electroluminescence occurs, then for a given electric field at the surface of the anode $E_a$, the distance at which electroluminescence begins, $r_T$, and the corresponding electroluminescence region length, $r_{EL}$, is \begin{equation}
    r_{EL} = r_T - r_a = r_a (\frac{E_a}{E_{min}}-1).
\end{equation} So for two different anodes with the same $E_a$, the thicker anode will have a larger electroluminescence region, and thus a larger gain. The trade-off is that a larger voltage has to be applied to the thicker wire to achieve the same $E_a$. Furthermore, the absolute $E_a$ does not seem to be the only factor in deciding whether or not an unmanageable level of light emission will occur. The results from the XeBrA experiment show that the breakdown electric field in liquid xenon is inversely related to the stressed electrode area (SEA) \cite{Watson:2022jvl}, i.e. the surface area of the electrode exposed to a high electric field, typically a significant fraction of the maximum field. Therefore, if the light emission is related to the breakdown of the electric field, then the dependence of the breakdown field on the SEA may dictate the optimal anode diameter.

\subsection{Potential for reactor neutrino CE$\nu$NS detection}
\label{sec:future_potential}

Coherent Elastic Neutrino Nucleus Scattering (CE$\nu$NS) is a standard model process with a considerable cross-section for large nuclei and low energies~\cite{COHERENT:2017ipa}. Nuclear reactors produce 6~$\bar{\nu}_e$ per fission \cite{Hayes:2016qnu}, there is, on average, 200~MeV per fission, which corresponds to roughly 2$\times10^{20}~\Bar{\nu}_e$/GJ. This has drawn considerable interest in recent years to place detectors near reactors to do searches for CE$\nu$NS, as well as new physics such as the neutrino magnetic moment and sterile neutrinos~\cite{Ni:2021mwa, Abdullah:2022zue}. However, CE$\nu$NS from reactor antineutrinos only produce sub-keV nuclear recoils in xenon, which is not energetic enough to produce detectable S1s. Therefore, CE$\nu$NS must be searched for by looking at few-electron S2 signals. Currently, the low $g_2$ value and the high light emission rate hinders the LXePSC's sensitivity to few-electron S2-only events.

\section{Conclusion}
In this work, we studied the performance of a liquid xenon proportional scintillation counter with a central thin (10~$\mu$m) diameter anode wire where electroluminescence is produced and detected by a surrounding array of PMTs. An ionization gain ($g_2$) of 1.6~PE/e$^-$, corresponding to an electroluminescence yield of $\sim$17 $\gamma$/e$^-$ is obtained with calibration events from $^{137}$Cs. The observed single electrons from photoemission of the cathode wires give an estimate for the SE gain of 1.8 PE/e$^-$, which is consistent with the ionization gain from the $^{137}$Cs calibration and indicates minor charge loss on electronegative impurities. Spurious light emission, unrelated to the single electrons, is observed and sets a limitation to the LXePSC performance. Low-energy electronic recoils from tritium beta decays are detected with similar efficiencies compared to the dual-phase LXeTPCs. Further increase of the electron gain is needed in order to improve the single electron resolution and detection efficiency for applications in detecting low energy neutrinos from a power reactor through the CE$\nu$NS process. Such a single-phase LXePSC with potentially suppressed single electron background may also find applications in light dark matter searches.

\acknowledgments

This research is sponsored by the US Defense Advanced Research Projects Agency (DARPA) under grant number HR00112010009, the content of the information does not necessarily reflect the position or the policy of the Government, and no official endorsement should be inferred. Jianyang Qi is supported by the High Energy Physics Consortium for Advanced Training (HEPCAT) graduate fellowship from the Department of Energy grant DE-SC0022313. Dr. Abigail Kopec is supported by the National Science Foundation Mathematical and Physical Sciences Ascending (MPS-Ascend) Postdoctoral Research Fellowship 2137911.

\printbibliography

@article{Wei:2021nuk,
    author = "Wei, Yuehuan and Qi, Jianyang and Shockley, Evan and Xu, Haiwen and Ni, Kaixuan",
    title = "{Performance of a radial time projection chamber with electroluminescence in liquid xenon}",
    eprint = "2111.09112",
    archivePrefix = "arXiv",
    primaryClass = "physics.ins-det",
    doi = "10.1088/1748-0221/17/02/C02002",
    journal = "JINST",
    volume = "17",
    number = "02",
    pages = "C02002",
    year = "2022"
}

@article{LUX:2020vbj,
    author = "Akerib, D. S. and others",
    collaboration = "LUX",
    title = "{Investigation of background electron emission in the LUX detector}",
    eprint = "2004.07791",
    archivePrefix = "arXiv",
    primaryClass = "physics.ins-det",
    doi = "10.1103/PhysRevD.102.092004",
    journal = "Phys. Rev. D",
    volume = "102",
    number = "9",
    pages = "092004",
    year = "2020"
}

@article{LZ:2022ufs,
    author = "Aalbers, J. and others",
    collaboration = "LZ",
    title = "{First Dark Matter Search Results from the LUX-ZEPLIN (LZ) Experiment}",
    eprint = "2207.03764",
    archivePrefix = "arXiv",
    primaryClass = "hep-ex",
    month = "7",
    year = "2022"
}

@article{Kopec:2021ccm,
    author = "Kopec, Abigail and Baxter, Amanda L. and Clark, Michael and Lang, Rafael F. and Li, Shengchao and Qin, Juehang and Singh, Riya",
    title = "{Correlated single- and few-electron backgrounds milliseconds after interactions in dual-phase liquid xenon time projection chambers}",
    eprint = "2103.05077",
    archivePrefix = "arXiv",
    primaryClass = "physics.ins-det",
    doi = "10.1088/1748-0221/16/07/P07014",
    journal = "JINST",
    volume = "16",
    number = "07",
    pages = "P07014",
    year = "2021"
}

@article{XENON:2022ltv,
    author = "Aprile, E. and others",
    collaboration = "XENON",
    title = "{Search for New Physics in Electronic Recoil Data from XENONnT}",
    eprint = "2207.11330",
    archivePrefix = "arXiv",
    primaryClass = "hep-ex",
    doi = "10.1103/PhysRevLett.129.161805",
    journal = "Phys. Rev. Lett.",
    volume = "129",
    number = "16",
    pages = "161805",
    year = "2022"
}

@article{XENONCollaborationSS:2021sgk,
    author = "Aprile, E. and others",
    collaboration = "(XENON Collaboration)\textsection{}, XENON",
    title = "{Emission of single and few electrons in XENON1T and limits on light dark matter}",
    eprint = "2112.12116",
    archivePrefix = "arXiv",
    primaryClass = "hep-ex",
    doi = "10.1103/PhysRevD.106.022001",
    journal = "Phys. Rev. D",
    volume = "106",
    number = "2",
    pages = "022001",
    year = "2022"
}

@article{Ni:2021mwa,
    author = "Ni, Kaixuan and Qi, Jianyang and Shockley, Evan and Wei, Yuehuan",
    title = "{Sensitivity of a Liquid Xenon Detector to Neutrino\textendash{}Nucleus Coherent Scattering and Neutrino Magnetic Moment from Reactor Neutrinos}",
    doi = "10.3390/universe7030054",
    journal = "Universe",
    volume = "7",
    number = "3",
    pages = "54",
    year = "2021"
}

@article{XENON:2019ykp,
    author = "Aprile, E. and others",
    collaboration = "XENON",
    title = "{XENON1T Dark Matter Data Analysis: Signal Reconstruction, Calibration and Event Selection}",
    eprint = "1906.04717",
    archivePrefix = "arXiv",
    primaryClass = "physics.ins-det",
    doi = "10.1103/PhysRevD.100.052014",
    journal = "Phys. Rev. D",
    volume = "100",
    number = "5",
    pages = "052014",
    year = "2019"
}

@article{Doke:2002oab,
    author = "Doke, Tadayoshi and Hitachi, Akira and Kikuchi, Jun and Masuda, Kimiaki and Okada, Hiroyuki and Shibamura, Eido",
    title = "{Absolute Scintillation Yields in Liquid Argon and Xenon for Various Particles}",
    doi = "10.1143/JJAP.41.1538",
    journal = "Jap. J. Appl. Phys.",
    volume = "41",
    pages = "1538--1545",
    year = "2002"
}

@article{XENON:2019izt,
    author = "Aprile, E. and others",
    collaboration = "XENON",
    title = "{XENON1T dark matter data analysis: Signal and background models and statistical inference}",
    eprint = "1902.11297",
    archivePrefix = "arXiv",
    primaryClass = "physics.ins-det",
    doi = "10.1103/PhysRevD.99.112009",
    journal = "Phys. Rev. D",
    volume = "99",
    number = "11",
    pages = "112009",
    year = "2019"
}

@article{Faham:2015kqa,
    author = "Faham, C. H. and Gehman, V. M. and Currie, A. and Dobi, A. and Sorensen, P. and Gaitskell, R. J.",
    title = "{Measurements of wavelength-dependent double photoelectron emission from single photons in VUV-sensitive photomultiplier tubes}",
    eprint = "1506.08748",
    archivePrefix = "arXiv",
    primaryClass = "physics.ins-det",
    doi = "10.1088/1748-0221/10/09/P09010",
    journal = "JINST",
    volume = "10",
    number = "09",
    pages = "P09010",
    year = "2015"
}

@article{Watson:2022jvl,
    author = "Watson, J. and others",
    title = "{Study of dielectric breakdown in liquid xenon with the XeBrA experiment}",
    eprint = "2206.07854",
    archivePrefix = "arXiv",
    primaryClass = "physics.ins-det",
    month = "6",
    year = "2022"
}

@article{Aprile:2014ila,
    author = "Aprile, E. and Contreras, H. and Goetzke, L. W. and Melgarejo Fernandez, A. J. and Messina, M. and Naganoma, J. and Plante, G. and Rizzo, A. and Shagin, P. and Wall, R.",
    title = "{Measurements of proportional scintillation and electron multiplication in liquid xenon using thin wires}",
    eprint = "1408.6206",
    archivePrefix = "arXiv",
    primaryClass = "physics.ins-det",
    doi = "10.1088/1748-0221/9/11/P11012",
    journal = "JINST",
    volume = "9",
    number = "11",
    pages = "P11012",
    year = "2014"
}

@article{Kuger:2021sxn,
    author = "Kuger, Fabian and Dierle, Julia and Fischer, Horst and Schumann, Marc and Toschi, Francesco",
    title = "{Prospects of charge signal analyses in liquid xenon TPCs with proportional scintillation in the liquid phase}",
    eprint = "2112.11844",
    archivePrefix = "arXiv",
    primaryClass = "physics.ins-det",
    doi = "10.1088/1748-0221/17/03/P03027",
    journal = "JINST",
    volume = "17",
    number = "03",
    pages = "P03027",
    year = "2022"
}

@article{Lin:2021izy,
    author = "Lin, Qing",
    title = "{Proposal of a Geiger-geometry single-phase liquid xenon Time Projection Chamber as potential detector technique for dark matter direct search}",
    eprint = "2102.06903",
    archivePrefix = "arXiv",
    primaryClass = "physics.ins-det",
    doi = "10.1088/1748-0221/16/08/P08011",
    journal = "JINST",
    volume = "16",
    number = "08",
    pages = "P08011",
    year = "2021"
}

@software{strax,
    title = {strax},
    url = {https://github.com/AxFoundation/strax/blob/master/strax/processing/peak\_splitting.py}}

@software{cylinterp,
    title = {cylinterp},
    url = {https://github.com/darkmatter-ucsd/cylinterp/tree/main}}

@article{GEANT4:2002zbu,
    author = "Agostinelli, S. and others",
    collaboration = "GEANT4",
    title = "{GEANT4--a simulation toolkit}",
    reportNumber = "SLAC-PUB-9350, FERMILAB-PUB-03-339, CERN-IT-2002-003",
    doi = "10.1016/S0168-9002(03)01368-8",
    journal = "Nucl. Instrum. Meth. A",
    volume = "506",
    pages = "250--303",
    year = "2003"
}

@article{COHERENT:2017ipa,
    author = "Akimov, D. and others",
    collaboration = "COHERENT",
    title = "{Observation of Coherent Elastic Neutrino-Nucleus Scattering}",
    eprint = "1708.01294",
    archivePrefix = "arXiv",
    primaryClass = "nucl-ex",
    doi = "10.1126/science.aao0990",
    journal = "Science",
    volume = "357",
    number = "6356",
    pages = "1123--1126",
    year = "2017"
}

@article{Hayes:2016qnu,
    author = "Hayes, A. C. and Vogel, Petr",
    title = "{Reactor Neutrino Spectra}",
    eprint = "1605.02047",
    archivePrefix = "arXiv",
    primaryClass = "hep-ph",
    doi = "10.1146/annurev-nucl-102115-044826",
    journal = "Ann. Rev. Nucl. Part. Sci.",
    volume = "66",
    pages = "219--244",
    year = "2016"
}

@article{Abdullah:2022zue,
    author = "Abdullah, M. and others",
    title = "{Coherent elastic neutrino-nucleus scattering: Terrestrial and astrophysical applications}",
    eprint = "2203.07361",
    archivePrefix = "arXiv",
    primaryClass = "hep-ph",
    month = "3",
    year = "2022"
}

@article{XENON100:2013wdu,
    author = "Aprile, E. and others",
    collaboration = "XENON100",
    title = "{Observation and applications of single-electron charge signals in the XENON100 experiment}",
    eprint = "1311.1088",
    archivePrefix = "arXiv",
    primaryClass = "physics.ins-det",
    doi = "10.1088/0954-3899/41/3/035201",
    journal = "J. Phys. G",
    volume = "41",
    pages = "035201",
    year = "2014"
}

@article{Bodnia:2021flk,
    author = "Bodnia, E. and others",
    title = "{The electric field dependence of single electron emission in the PIXeY two-phase xenon detector}",
    eprint = "2101.03686",
    archivePrefix = "arXiv",
    primaryClass = "physics.ins-det",
    doi = "10.1088/1748-0221/16/12/P12015",
    journal = "JINST",
    volume = "16",
    number = "12",
    pages = "P12015",
    year = "2021"
}

@article{Szydagis:2022ikv,
    author = "Szydagis, M. and others",
    title = "{A Review of NEST Models, and Their Application to Improvement of Particle Identification in Liquid Xenon Experiments}",
    eprint = "2211.10726",
    archivePrefix = "arXiv",
    primaryClass = "hep-ex",
    month = "11",
    year = "2022"
}

@article{Park:1994hd,
    author = "Park, J. and Atac, M. and Cline, D. B. and Wang, H. and Smith, P. F.",
    editor = "Cline, D. B.",
    title = "{Dark Matter and Neutrino Detection with Liquid Xenon}",
    journal = "Conf. Proc. C",
    volume = "940216",
    pages = "288--295",
    year = "1994"
}

@article{Masuda:1978tjp,
    author = "Masuda, K. and Takasu, S. and Doke, T. and Takahashi, T. and Nakamoto, A. and Kubota, S. and Shibamura, E.",
    title = "{A LIQUID XENON PROPORTIONAL SCINTILLATION COUNTER}",
    reportNumber = "WU-HEP-78-5",
    doi = "10.1016/0029-554X(79)90600-1",
    journal = "Nucl. Instrum. Meth.",
    volume = "160",
    pages = "247--253",
    year = "1979"
}

@mastersthesis{Wang:1998gq,
    author = "Wang, Hanguo",
    title = "{WIMP Detection using Liquid Xenon}",
    reportNumber = "UMI-99-06762",
    type = "{Ph.D. thesis, University of California Los Angeles}",
    year = "1998"
}

@software{szydagis_m_2018_1314669,
  author       = {Szydagis, M. and
                  Balajthy, J. and
                  Brodsky, J. and
                  Cutter, J. and
                  Huang, J. and
                  Kozlova, E. and
                  Lenardo, B. and
                  Manalaysay, A. and
                  McKinsey, D. and
                  Mooney, M. and
                  Mueller, J. and
                  Ni, K. and
                  Rischbieter, G. and
                  Tripathi, M. and
                  Tunnell, C. and
                  Velan, V. and
                  Zhao, Z.},
  title        = {Noble Element Simulation Technique v2.0},
  month        = jul,
  year         = 2018,
  publisher    = {Zenodo},
  version      = {v2.0.0},
  doi          = {10.5281/zenodo.1314669},
  url          = {https://doi.org/10.5281/zenodo.1314669}
}

\end{document}